\documentstyle[prb,aps,amsfonts,amssymb,floats,epsfig]{revtex}
\begin{document}
\draft 
\twocolumn[\hsize\textwidth\columnwidth\hsize\csname @twocolumnfalse\endcsname
\title{Anisotropy of the optical conductivity of high T$_c$ cuprates} 
\author{D. van der Marel}
\address{Laboratory of Solid State Physics, Materials Science Centre, 
Nijenborgh 4, 9747 AG Groningen, The Netherlands} 
\date{\today}
\maketitle
\begin{abstract}
The optical conductivity along ($\sigma_{xx}$) and perpendicular 
($\sigma_{zz}$) to the planes is calculated assuming strong 
$k_{\parallel}$-dependence of the scattering rate and the     
c-axis hopping parameter. 
The closed analytical expressions for $\sigma_{xx}(\omega)$ 
and  $\sigma_{zz}(\omega)$ are shown to be integrable at low and high frequencies. 
A large and qualitatively different frequency dependence for both polarizations
follows directly from the model. 
The expression for $\sigma_{xx}(\omega)$  has an effective scattering rate proportional to 
frequency, and can be easily generalized to provide a simple analytical expression, which 
may replace the Drude formula in the case of  non-Fermi liquids.
\end{abstract}
\pacs{74.72.-h,74.25.Gz}
\vskip2pc]
\begin{figure}
\centerline{\psfig{figure=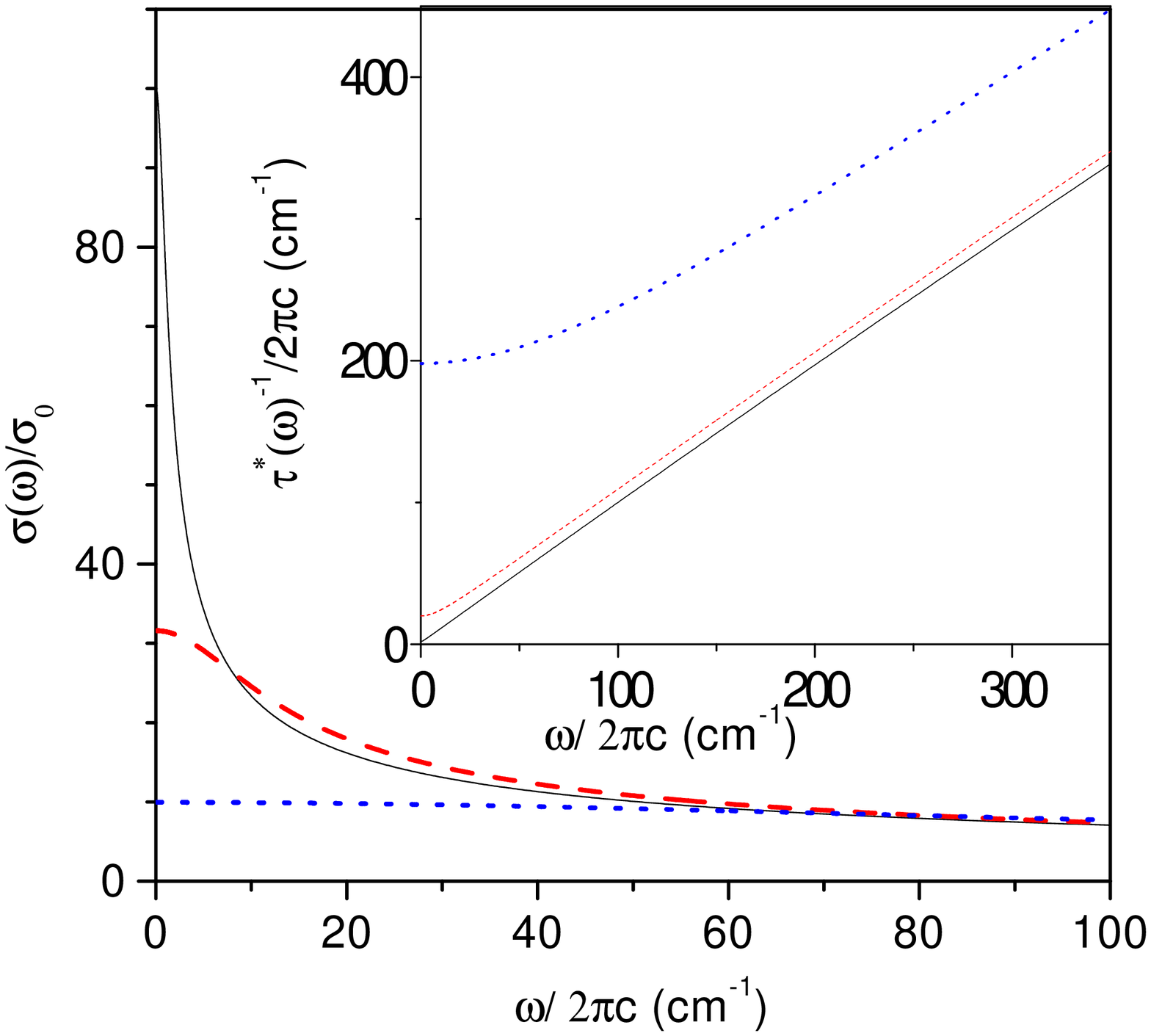,width=7cm,clip=}}
  \caption{Optical conductivity, $\sigma_{xx}$, using Eq. 3, 
   with parameters $\hbar \Gamma = 1.24 eV$ and 
   $\hbar/\tau = 0.124 meV$ (solid curves, T = 14 K ),
   $\hbar/\tau = 1.24 meV$ (dashed curves, T = 44 K ), 
   $\hbar/\tau = 12.4 meV$ (dotted curves, T = 140 K). The temperatures
   correspond to $1/\tau=T^2/T_0$, adopting the value $T_0 = 12 meV$ from
   Ioffe and Millis.
   Inset: Frequency dependent scattering corresponding to the conductivities displayed in the
   main panel. }                                                  
 \label{comxx}
\end{figure}

\begin{figure}
\centerline{\psfig{figure=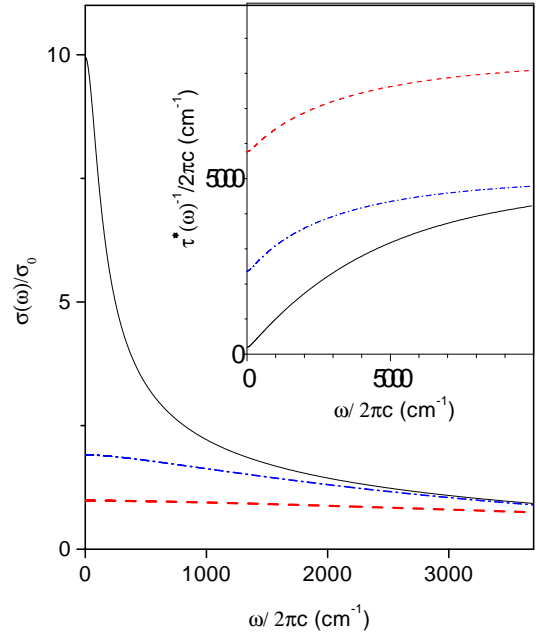,width=7cm,clip=}}
  \caption{Optical conductivity using the parameters
  $\hbar/\tau = 12.4 meV$ and $\hbar\Gamma = 1.24 eV$.
  Solid curves $\sigma_{xx}$, using Eq. 3. 
  Dashed: $\sigma_{zz}$ for the simple tetragonal structure using Eq. 11.
  Dashed-dotted: $\sigma_{zz}$ for the body centered tetragonal structure
  using Eq. 14.
  Inset: Frequency dependent scattering corresponding to the conductivities displayed in the
  main panel.}                                                  
 \label{siggam}
\end{figure}

\begin{figure}
\centerline{\psfig{figure=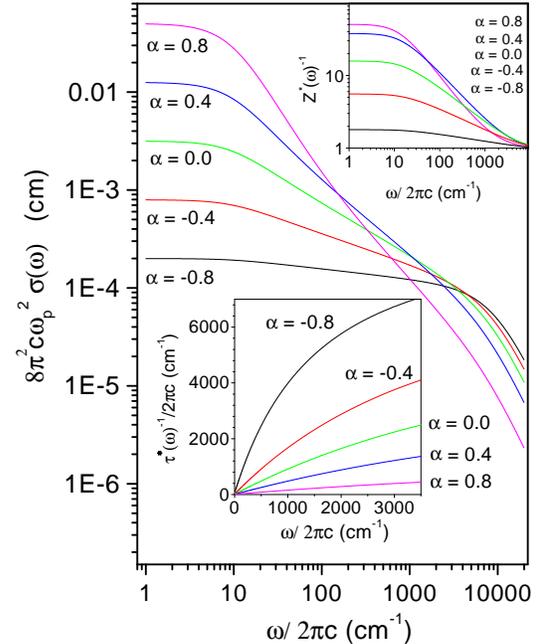,width=7cm,clip=}}
 \caption{Optical conductivity using Eq. 15, with the parameters
  $\hbar/\tau = 1.24 meV$ and $\hbar\Gamma = 1.24 eV$, and
  $\alpha = 0.8, 0.4, 0.0, -0.4, \mbox{and} -0.8$
  Insets: Frequency dependent scattering rate and spectral weight factor.}                                                  
 \label{nfl}
\end{figure}

\narrowtext
The anomalous transport characteristics in the normal state 
of the high T$_c$ superconductors\cite{varma,pwa}, and the corresponding 
quasi-particle properties in the superconducting state have been discussed in a series
of recent papers emphasizing strong anisotropy of quasiparticles as a function of position
around the Fermi-surface\cite{xiang,pines,millis}. Such anisotropies have been used 
both in the framework
of non-Fermi-liquid models {\em and} Fermi-liquid motivated models\cite{chakravarty,sudbo}. 
In the models considered in Refs.\cite{pines,millis},
anomalous strong scattering of quasi-particles is assumed in the Fermi surface regions 
closest to the saddle-points, while  weak scattering is assumed in the zone-diagonal 
directions in 
$k$-space along $0$ to $(\pi,\pi)$. Stojkovic and Pines\cite{pines} used a frequency and k-dependent
self-energy for numerical calculations of the in-plane optical properties and
magnetotransport of the cuprates. Ioffe and Millis (I\&M)\cite{millis} proposed an analytic expression
for the $k$-dependent self energy, containing {\em no} explicit frequency dependence. 
The resulting expression for the optical conductivity corresponds to a Fermi-surface
average of Lorentzian response functions with a $k$-dependent scattering rate. Interestingly
this expression may be 'correct' (at least approximately) irrespective whether
the underlying model is Fermi-liquid or not. The strong momentum depedence of the
scattering rate is experimentally motivated by Angle Resolved Photo Electron
Spectra\cite{shen}. We will see in this paper, that this model
provides an excellent description of both the out-of-plane and in-plane optical
properties of high T$_c$ superconductors. We take the expression of Ref.\ \cite{millis}
for the Fermi-surface averaged optical conductivity as our starting point
\begin{equation}
  \sigma_{xx}(\omega) = \frac{n e^2}{2\pi m_{xx}}\int_0^{2\pi}d\theta 
  \frac{i}{\omega+i\gamma(\theta)}
 \label{sxx1}
\end{equation}
In this expression $\theta$ is the angle along the Fermi surface relative to the $(\pi,\pi)$
direction, $n=k_F^2/2\pi d$ is the density of electrons, $m_{xx}=\hbar k_F / v_F$ is the 
planar effective mass, and $d$ the distance between planes. For $\gamma(\theta)$
isotropic the classical Drude formula is recovered. (Note a factor $1/4\hbar d$ difference 
with Eq. 13 of Ref.\ \cite{millis}). I\&M furthermore postulated the following formula for the 
k-dependent scattering rate, motivated by photoemission experiments on a variety of high T$_c$ 
superconductors
\begin{equation}
  \gamma(k_{\parallel}) = 1/\tau + \Gamma\sin^{2}(2\theta) 
  \label{gtet} 
\end{equation}
Inserting  this expression for the scattering rate in Eq. \ref{sxx1}, substituting 
$\xi=e^{2i\theta}$, carrying out the corresponding contour integral, and
using Cauchi's principle value method, we obtain the planar conductivity
\begin{equation}
  \sigma_{xx}(\omega) =  
    \frac{\sigma_{xx,0} }{\sqrt{1-i\omega\tau}\sqrt{1+\Gamma\tau-i\omega\tau}} 
 \label{sxx2}
\end{equation}
where $\sigma_{xx,0}\equiv ne^2\tau/m_{xx}$.
In the approximation of Ref.\ \cite{millis}, valid close to the zone-diagonals, the term 
$(\omega+i\Gamma)^{1/2}$ is replaced with 
the constant $(i\Gamma)^{1/2}$, resulting in non-integrable behavior of 
$\mbox{Re}\sigma_{xx}(\omega)$ at high frequencies.
I\&M adopted the model assumption, that $1/\tau$ 
should have a quadratic temperature dependence, $1/\tau=T^2/T_0$, 
as in a classical Fermi-liquid. (But 
note, that {\em only} the zone-diagonal directions have a scattering rate which becomes zero
at low temperature in their model, which is anything but Fermi-liquid 
behavior). The DC-resistivity then has the linear temperature dependence
$\rho=\Gamma^{1/2}T_0^{-1/2}m_{xx}T/(ne^2)$,
observed in optimally doped cuprates.
\\
Let us briefly investigate some of the main analytical properties of the above result. First
we notice, that $\sigma(\omega)=\sigma(-\omega)^*$ is satisfied. Let us now consider the
frequency dependent scattering rate, using the definition 
$1/\tau^*(\omega)\equiv\omega\mbox{Re}\sigma/\mbox{Im}\sigma$,
commonly used for the
analysis of optical spectra of high T$_c$ superconductors and heavy Fermion systems 
\begin{equation}
 \frac{1}{\tau^*(\omega)}
    =
 \frac{
       \left(1+\Gamma\tau\right)
       \sqrt{1+\left(\omega\tau\right)^2} 
          +
       \sqrt{\left(1+\Gamma\tau\right)^2+\left(\omega\tau\right)^2}
      } 
      {\tau\left[
       \sqrt{1+\left(\omega\tau\right)^2}
       +
       \sqrt{\left(1+\Gamma\tau\right)^2+\left(\omega\tau\right)^2}
       \right]
       }                                         
\end{equation}
At intermediate frequencies, $1/\tau \ll \omega \ll \Gamma$, the scattering rate
$1/\tau^*(\omega)=\omega$. 
The high frequency limiting behavior of these quantities gives
\begin{equation}
 \lim_{\omega\rightarrow\infty} \sigma_{xx}(\omega)  =  
 \frac{n e^2}{m_{xx}}\frac{i}{\omega+i(1/\tau+\Gamma/2)}
 \label{hisxx}
\end{equation}
The low frequency limiting behavior corresponds to a Drude conductivity 
\begin{equation}
 \sigma_{D}(\omega) = \lim_{\omega\rightarrow 0} \sigma_{xx}(\omega) = 
 \frac{Z^*(0) n e^2}{m_{xx}}\frac{i}{\omega+i/\tau^*(0)}
 \label{losxx}
\end{equation}                              
with the scattering rate $\tau^*(0)$ and the spectral weight $Z^*(0)$ defined as
\begin{displaymath}
\tau^*(0)=\frac{\tau}{2}\frac{\Gamma\tau+2}
{\Gamma\tau+1}=\frac{T_0}{2T^2}+\frac{1}{2\Gamma} + ..., 
\end{displaymath}
\begin{equation}
Z^*(0)=\frac{\int_0^{\infty}\sigma_{D}(\omega) d\omega}
{\int_0^{\infty} \sigma_{xx}(\omega) d\omega} 
= \frac{2}{\sqrt{1+\Gamma\tau}}
= \frac{2T}{\sqrt{\Gamma T_0} } - \frac{T^3}{\{\Gamma T_0\}^{3/2}} +...
\end{equation}
where the series expansion is relevant for at low temperatures, where
$\tau$ diverges. 
Taken together, we see that $\sigma_{xx}$ is integrable, and satisfies
\begin{equation}
 \int_0 ^{\infty} \mbox{Re} \sigma_{xx}(\omega) d\omega =  \frac{\pi n e^2}{2m_{xx}}
 \label{sum1}                                                
\end{equation}
\\ 
We see, that for $T\ll \sqrt{T_0\Gamma}$, which is of the order of 1000 K, 
$1/\tau^*(0)$ is proportional to $T^2$, and that the low frequency 
spectral weight $Z^*(0)$ increases linearly with temperature. It is amusing, that, apart 
from a factor two, the inverse Hall constant 
derived in Ref.\cite{millis} is precisely this spectral weight: 
$1/R_H=B\sigma_{xx}^2/\sigma_{xy}=2e Z^*(0)n$.
\\
In various papers a linear temperature dependence of the scattering rate of
the low-frequency Drude tail has been reported. However, a fit to Hagen-Rubens behavior
provides the DC conductivity $\sigma_{xx}=Z^*(0)ne^2\tau^*(0)/m_{xx}$. Usually
a temperature {\em in}dependent Drude spectral weight is imposed 
for the purpose of minimizing the number of fit parameters. However, the
'Drude scattering rate' resulting from such a fit, is not a faithful 
representation of the actual line-width of $\sigma_D(\omega)$ if
$Z^*(0) < 1$. Instead one can employ the fact, that
$1/\tau^*(\omega) = \omega\mbox{Re}\sigma_{xx}(\omega)/\mbox{Im}\sigma_{xx}(\omega)$
and $Z^*(\omega) =  \mbox{Im}[-ne^2/(4m_{xx}\pi\omega\sigma_{xx}(\omega)]$ can be
obtained from a careful analysis of the DC limiting behavior of the experimentally
measured real and imaginary part of the conductivity. This 
should reveal both the quadratic temperature dependence of $1/\tau$ and the linear
temperature dependence of the low frequency spectral weight. Although this behavior has, to my best knowledge, 
not been mentioned explicitly
in the literature, it may be contained in the experimental infrared data reported by
various groups\cite{tanner}. This aspect certainly requires further scrutiny, and calls
for high precision optical experiments in the far infrared range. 
\\
In Fig.\ \ref{comxx} the evolution of $\sigma_{xx}(\omega)$ with temperature is displayed, mimicked
by varying the parameter $1/\tau$. This behavior is identical to what has been  
observed experimentally for the in-plane optical conductivity \cite{artem,tanner,somal}. 
Note also, that the total spectral weight under the optical conductivity curves is 
the same for all values of $1/\tau$.
\\
Let us now investigate the behavior of $\sigma_{zz}$.
Recently Xiang and Wheatley\cite{xiang} calculated the c-axis penetration depth assuming 
d-wave pairing, and assuming a model for the c-axis transport where
momentum parallel to the planes, $k_{\parallel}$, is conserved. Based 
on LDA band theoretical results \cite{OKA} they argued, that  
for HTSC's with a simple (non-body centered) tetragonal
structure, $t_{\perp} \propto (\cos(k_x a) - \cos(k_y a))^{2}$, 
leading to a $T^5$ powerlaw at low temperature. Following the
approach used in Refs. \cite{xiang,pines,millis} 
this $k$-dependence is evaluated as a function of $k_{\parallel}$
around the Fermi surface. 
\\
The c-axis conductivity is, in leading orders of $t_{\perp}$ \cite{sigzz}
\begin{equation}
 \sigma_{zz}(\omega) =\frac{e^2 d m_{xx}}{\pi^2 \hbar^4} \int_0^{2\pi} d\theta 
  \frac{t_{\perp}(\theta)^2}{-i\omega+\gamma(\theta)}
  \label{szz1}
\end{equation}                 
A simple hole, or electron-type surface is circular, and
has the following $k_{\parallel}$-dependence in leading orders of the harmonic expansion
\begin{equation}
t_{\perp}(k_{\parallel}) = t_0 \sin^{2}(2\theta) 
\end{equation}                 
Note, that the expression for $\sigma_{zz}$ now contains a weighting factor proportional
to $\sin^4(2\theta)$, which is sharply peaked along $(0,\pm\pi)$ and $(\pm\pi,0)$. Again
the integral can be solved using Cauchi's principle value method, 
with the result
\begin{equation}
 \frac{\sigma_{zz}(\omega)}{\sigma_{zz,0}}=1+
 2\Omega^2\left[\sqrt{\frac{\Omega^2}{1+\Omega^2}} - 1\right] 
 \label{szzstet}
\end{equation}
where $\Omega \equiv \sqrt{(1-i\omega\tau)/(\Gamma\tau)}$,
and $\sigma_{zz,0} \equiv e^{2}d m_{xx}t_0^2/(\pi\hbar^{4}\Gamma)$.
I\&M needed a value of $\hbar \Gamma$ of about 0.15 eV to fit 
$\sigma_{xx}(\omega)$ of YBCO\cite{joe}. In fact, if we use Eq.\ref{sxx2}
for $\sigma_{xx}(\omega)$, a larger value is required, 
$\hbar \Gamma \approx 1.2 eV$ to produce a good fit. Both values place the c-axis-transport well 
inside the dirty limit region, if we adopt Eq. \ref{szzstet} as an
expression representative of the c-axis optical conductivity in the normal state of  
high T$_c$ superconductors. 
\\
The analytical formula for $1/\tau^*(\omega)$ according
to Eq. \ref{szzstet} is a rather lengthy expression, which is not very illuminating. 
To obtain further insight in 
the effect of a strongly anisotropic scattering rate around the Fermi surface on
the optical properties, in Fig.\ \ref{siggam} both $\sigma_{xx}$ and $\sigma_{zz}$ and the
corresponding frequency dependent scattering rates are presented,
adopting $\hbar/\tau =12.4 meV$, and $\hbar\Gamma = 1.24 eV$ as model
parameters. Clearly $\sigma_{zz}$ is dominated by the transport in the regions close to
the saddlepoints, resulting in a frequency dependent scattering rate ranging from about 
$\Gamma/2$ at low frequencies, to $\Gamma$ at high frequencies. The calculated
spectral shape resembles closely the experimental data
\cite{cooper,juergen,tamasaku,kim,homes,basov,hosseini,markus,dulic}.
\\
At this point it should be noted however, that the model assumption for the 
$k_{\parallel}$-dependence  of $t_{\perp}$ is only justified for cuprates with
the simple tetragonal structure, in particular Hg1201 or Tl1201. 
Materials like LSCO, Tl2201, and
Bi2212, which have been  studied more intensively during recent years, have a
body centered tetragonal structure, for which
\begin{equation}
t_{\perp} \propto (\cos(k_x a) - \cos(k_y a))^{2} \cos(k_x a/2) \cos(k_y a/2)
\end{equation}
Hole-type Fermi-surfaces cross not only the zone-diagonal directions, but also
the lines $k_x = \pi$ and $k_y = \pi$, where extra zeros occur. The integration over $\theta$
in Eq. \ref{szz1} thus contains eight zero's, along the $(0,\pm\pi)$, $(\pm\pi,0)$, 
and $(\pm\pi,\pm\pi)$ directions.
The lowest order harmonic expansion of the angular dependence in this case is
\begin{equation}
t_{\perp}(k_{\parallel}) = t_0 \sin^{2}(4\theta) 
\end{equation}                 
resulting in the following expression for the c-axis optical conductivity
\begin{equation}
  \frac{\sigma_{zz}(\omega)}{2\sigma_{zz,0}} = 
   (1+2\Omega^2)(1-8\Omega^4-8\Omega^2) + 16\Omega^3(1+\Omega^2)^{3/2} 
  \label{szzbctet}
\end{equation}
The real part of this expression, and the corresponding $1/\tau^*(\omega)$ 
are displayed in Fig.\ \ref{siggam}, using the same parameters as before. We
observe from this Figure, that for the body centered sytems $\sigma_{zz}(\omega)$
is somewhat more peaked. This results from the fact that contributions 
from the regions close to the saddle points are suppressed in this case.\\
\\                               
Eq.\ \ref{sxx1} for $\sigma_{xx}$, suggests a form wich is of general
use in other materials (not necessarily 2-dimensional) to analyze the optical conductivity of 
non-Fermi liquids with a simple analytical formula
\begin{equation}
  \sigma_{A}(\omega) = 
  \frac{\omega_p^2}{4\pi}\frac{i}{(\omega+i/\tau)^{1-2\alpha}(\omega+i\Gamma)^{2\alpha}}
 \label{sa}
\end{equation}
If we set $\tau\rightarrow\infty$, and consider only $\omega \ll \Gamma$, this corresponds 
to the expresion derived by Anderson\cite{pwa} based on the Luttinger liquid model 
for high T$_c$ superconductors. The notation for
the coefficient $\alpha$ was adopted accordingly. In Anderson's paper Fermi-liquid
response is restored for $\alpha\rightarrow 0$, as in the present case. 
Let us briefly investigate some of the main analytical properties of $\sigma_{A}$. 
Again $\sigma_{A}(\omega)=\sigma_{A}(-\omega)^*$ is satisfied. 
The high frequency limiting behavior of $\sigma_{A}$ 
\begin{equation}
 \lim_{\omega\rightarrow\infty} \sigma_{A}(\omega)  =  
\frac{\omega_p^2}{4\pi}\frac{1}{(1-2\alpha)/\tau+2\alpha\Gamma-i\omega}
\end{equation}
garantees, that $\sigma_{A}$ is integrable, and satisfies
\begin{equation}
 \int_0^{\infty}  \mbox{Re} \sigma_{A}(\omega) d\omega = \frac{1}{8}\omega_p^2
\end{equation}
At the low frequency end
\begin{equation}
 \lim_{\omega\rightarrow 0} \sigma_{A}(\omega) =  
 \frac{\omega_p^2}{4\pi}
 \frac{\tau^{(1-2\alpha)}\Gamma^{-2\alpha}}
 {1-i\omega[(1-2\alpha)\tau+2\alpha/\Gamma]}
 \label{sum2}
\end{equation}
which corresponds to a narrow Drude peak with an effective
carrier life-time $\tau^*(0)=(1-2\alpha)\tau+2\alpha/\Gamma$. 
The frequency dependent scattering rate  $1/\tau^*(\omega)$ crosses over 
to a linear frequency dependence
for $1/\tau \ll \omega \ll \Gamma$, and acquires a constant value 
$(1-2\alpha)/\tau+2\alpha\Gamma$ for $\omega \rightarrow \infty$. Hence the
linear frequency dependence of the scattering rate is a robust property of the above 
phenomenological expression for $\sigma_{A}$. To illustrate this point, in Fig.\ \ref{nfl}
$\sigma_{A}(\omega)$, the corresponding $1/\tau^*(\omega)$ and $1/Z^*(\omega)$ are 
displayed for a number of different values of $\alpha$.
The Luttinger liquid
analysis of experimental spectra\cite{pwa,bontemps} gave $\alpha=0.15\pm0.05$. The analysis of
Ioffe and Millis corresponds to $\alpha=0.25$, which is reasonably close.
\\
In conclusion, the combination of the k-dependent scattering rate, 
and the k-dependence of the inter-plane hopping parameter 
known from band theorie, leads to a quantitative
description of the strongly anisotropic optical properties found in the high T$_c$ cuprates.
Both the anomalous in-plane optical conductivity {\em and} the c-axis conductivity follow
from the assumption of the k-dependent scattering rate, which has been
recentely proposed by Stojkovich, Pines\cite{pines}, Ioffe and Millis\cite{millis}. 
Closed analytical expressions were
obtained for the optical conductivity, which are really quite simple, and were
further extended and discussed in a phenomenological framework of non-Fermi liquids.
The present result corresponds closely to experimental data of $\sigma_{xx}(\omega)$
\cite{joe,schlesinger,thomas,bontemps}, including the high frequency cutoff and the low frequency 
cross-over to Drude behavior as can be seen from the saturation, and of $\sigma_{zz}(\omega)$
\cite{juergen,tamasaku,kim,homes,basov}. 
\\                                                             
{\bf Acknowledgements}
It is a pleasure to acknowledge stimulating discussions with W. N. Hardy and D. Dulic, and
useful comments on the manuscript by H. J. Molegraaf, and M. Gr\"uninger.
\end{document}